\documentclass[aps,prb,superscriptaddress,showpacs,twocolumn]{revtex4}
\usepackage{graphicx}
\begin{document}
\title{Nonlocal vortex motion in mesoscopic amorphous Nb$_{0.7}$Ge$_{0.3}$ structures}
\author{A.~Helzel}
\affiliation{Institute for Experimental and Applied Physics,
University of Regensburg, D-93025 Regensburg, Germany}
\author{I.~Kokanovi\'{c}}
\affiliation{Institute for Experimental and Applied Physics,
University of Regensburg, D-93025 Regensburg,
Germany}\affiliation{Department of Physics, Faculty of Science,
University of Zagreb, Bijeni\v{c}ka 32, HR-10000 Zagreb, Croatia}
\author{D.~Babi\'{c}}
\altaffiliation[Corresponding author] {} \email{dbabic@phy.hr}
\affiliation{Department of Physics, Faculty of Science, University
of Zagreb, Bijeni\v{c}ka 32, HR-10000 Zagreb, Croatia}

\author{L.~V.~Litvin}
\affiliation{Institute for Experimental and Applied Physics,
University of Regensburg, D-93025 Regensburg, Germany}

\author{F.~Rohlfing}
\affiliation{Institute for Experimental and Applied Physics,
University of Regensburg, D-93025 Regensburg,
Germany}
\author{F.~Otto} \affiliation{Institute for Experimental and Applied Physics,
University of Regensburg, D-93025 Regensburg, Germany}

\author{C.~S\"{u}rgers}
\affiliation{Physikalisches Institut and DFG Center for Functional Nanostructures (CFN),
Universit\"{a}t Karlsruhe, D-76128 Karlsruhe, Germany}
\author{C.~Strunk}
\affiliation{Institute for Experimental and Applied Physics,
University of Regensburg, D-93025 Regensburg, Germany}%
%
%
%
%
\begin{abstract}
We study nonlocal vortex transport in mesoscopic amorphous
Nb$_{0.7}$Ge$_{0.3}$ samples. A dc current $I$ is passed through a
wire connected via a perpendicular channel, of a length $L= 2-5$
$\mu$m, with a pair of voltage probes where a nonlocal response
$V_{nl} \propto I$ is measured. The maximum of $R_{nl}=V_{nl}/I$ for
a given temperature occurs at an $L$-independent magnetic field and
is proportional to $1/L$. The results are interpreted in terms of
the dissipative vortex motion along the channel driven by a remote
current, and can be understood in terms of a simple model.

\end{abstract}
%
%
%
%

%
\maketitle

In a pioneering work Giaver measured a magnetic-flux-transformer
effect in type II superconductors.\cite{giaver} He applied a
magnetic field $B$ perpendicularly to a sample comprising two
superconducting sheets separated by a thin insulator, passed a
current $I$ through one of the superconductors and measured a
voltage developed over the other one - where no current was flowing.
The induced voltage was a consequence of an electromagnetic coupling
of vortices in the two layers. In their recent experiment Grigorieva
{\it et al.}\cite{grigorieva} demonstrated a complementary
flux-transformer phenomenon associated with vortices. They produced
mesoscopic amorphous MoGe structures of a double-cross shape,
consisting of two parallel wires connected at a right angle by a
channel of a width $w= 0.07 - 2$ $\mu$m and a length $L = 0.5 - 12$
$\mu$m. In a perpendicular $B$ and with $I$ through one of the
parallel wires a nonlocal voltage $V_{nl}$ appeared over the second,
current-free wire. This novel, transversal flux-transformer effect
originated in the in-plane vortex-vortex repulsion which
conveyed the driving force from the current-carrying wire to the
vortices in the channel. The effect disappeared not only for $L$
exceeding 6-7~$\mu$m but also for $w$ larger than $\sim 0.5 -
1$~$\mu$m. When $w$ was sufficiently small the force on the vortices
in the channel was transferred over many intervortex distances, and,
moreover, $V_{nl}$ was proportional to $I$. The efficiency of 
transversal flux-transformer effect
can be quantified by a nonlocal resistance $R_{nl}=V_{nl}/I$.

In the experiment of Grigorieva {\it et al.}\cite{grigorieva} the
local mixed-state dissipation was characterised on separate mm-sized
films, whereas $V_{nl}$ was measured by a low-frequency ac method
during $B$ sweeps at constant temperatures $T$. An ac method was
used because $V_{nl}$ was in nV range, i.e. $R_{nl} < 5$~m$\Omega$,
thus being too small for a dc detection. In our work we focused on
dc probing of transversal flux-transformer effect 
and measuring $V_{nl}$ and the local voltage
$V_l$ on a same sample, which was possible in multi-terminal
amorphous ($a$-)Nb$_{0.7}$Ge$_{0.3}$ structures of the geometry shown
in the inset to Fig.~1. The weak pinning, characteristic of the
$a$-Nb$_{0.7}$Ge$_{0.3}$ material used, resulted in a dc-measurable
$V_{nl}$ and $R_{nl} \sim 1\;\Omega$ even at very low temperatures.
The measured nonlocal resistance was hence two orders of magnitude larger
than in Ref.~\onlinecite{grigorieva}.

We measured the transversal flux-transformer effect in samples of
different length by isothermal sweeps of $B$ at different applied
$I$ and for $0.15\,T_c \lesssim T \lesssim 0.95 \,T_c$, where $T_c$
is the superconducting transition temperature. $V_{nl}$ depends
linearly on $I$ in the range $I = 0.1 - 1$ $\mu$A. With increasing
$B$, $R_{nl}(B)$ first acquires a nonzero value at a certain
magnetic field $B_d$, then has a maximum at some $B_p$ and gradually
vanishes close to the upper critical magnetic field $B_{c2}$. The
main representatives of $R_{nl}$, i.e. $B_p$ and $R_p =
R_{nl}(B_p)$, behave differently with respect to the channel length.
$B_p$ is independent of $L$ whereas $R_p \propto 1/L$, suggesting a
 vortex velocity $u_{nl} \propto 1/L$ at the non-local
voltage probes. This we relate to the total frictional force on the
vortices in the channel being proportional to $L$.

We investigated two structures of the type shown in the inset to
Fig.~1, where we also assign numbers to the leads and define the
coordinate system (with the unit vectors ${\bf \hat{x}}$, ${\bf
\hat{y}}$ and ${\bf \hat{z}}$). As in our previous
studies,\cite{holes,weffi} the samples were produced by combining
electron-beam lithography with magnetron sputtering but the film
thickness was increased from $20$ nm to $d=60$ nm in order to safely
avoid inhomogeneities detected in samples of a cross section smaller
than $\sim 5000$ nm$^2$. This way the pinning was enhanced, as we
inferred from the local dissipation, but still remained weak enough
to permit a nonlocal vortex motion over several microns. Contacts 0
and 1 are used for applying a dc $I=I_{01}$ through the horizontal
wire, exerting a local force on vortices in the $y$ direction (since
${\bf B}= -B {\bf \hat{z}}$). Having a velocity ${\bf u} = u {\bf
\hat{y}}$, vortices induce an electric field ${\bf E} = E {\bf
\hat{x}}$. Combinations of contacts $i,j = 6-9$ are used for
measuring the local voltage drop $V_l$. The local vortex pressure is
transferred from the horizontal wire along the zero-$I$ channels
(the length $L$ is indicated for the left channel) toward the
crosses contacted by leads 2,3 and 4,5, where $V_{nl}$ is measured.
For the first sample the channel lengths were 2 and 3~$\mu$m, and
for the second 3 and 5~$\mu$m, so we covered the range
$L=2-5$~$\mu$m and had two $L=3$~$\mu$m samples for consistency
check. All wires of a single sample had the same width, $w=275$~nm
for the sample with the longer channels and $w=250$~nm for the other
one. This small difference did not affect the results presented
henceforth. Since dc measurements are invariably burdened by
sub-$\mu$V parasitic signals, in order to determine $V_{nl}(I)$ for
a given measurement with leads $i,j$ we recorded both $V_{nl,ij}(I)$
and $V_{nl,ij}(I=0)$ taken at the same $B$-sweep rate and direction.
$V_{nl,ij}(I=0)$ was different for different pairs of leads and
depended very weakly on the $B$-sweep rate. The background-free
nonlocal voltage was extracted as $V_{nl}(I) = V_{nl,ij}(I) -
V_{nl,ij}(I=0)$. By this procedure we found a very regular behaviour
of $V_{nl}(I)$, which, in particular, for the two $L=3$~$\mu$m
samples agreed within the error bars. From the local voltage
response between contacts (8,9) we characterised the samples in
the same way as in our previous work,\cite{weffi} obtaining
$T_c=2.95$ K, the normal-state resistivity $\rho_n= 3.7$ $\mu
\Omega$m, $-(dB_{c2}/dT)_{T=T_c} = 2.17$ T/K, and the
Ginzburg-Landau parameters $\kappa = 100$, $\xi (0)= 7.15$ nm and
$\lambda (0) = 1.18$ $\mu$m.

\begin{figure}
\includegraphics[width=75mm]{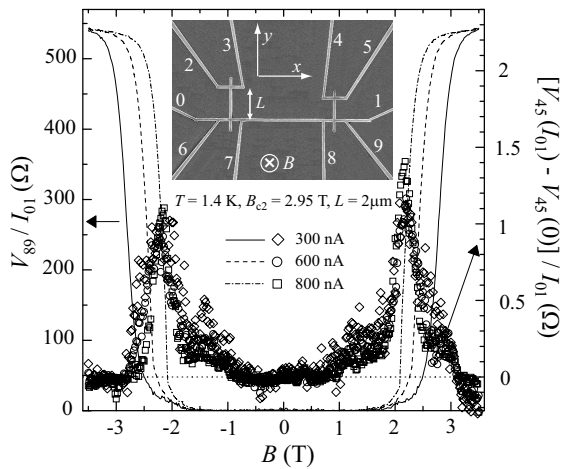}
\caption{Local (lines, left-hand scale) and nonlocal (symbols, right-hand scale)
response for the $L = 2$ $\mu$m
channel at $T=1.4$ K ($B_{c2}=2.95$ T) and $I=I_{01}=300, 600, 800$ nA.
Inset: A photograph of the sample, with the designation of the
leads and the definition of the coordinate system.}
\end{figure}

Results typical of $V_{nl}$ are shown in Figs.~1 and 2. The data in
Fig.~1 were obtained for the $L=2$~$\mu$m  channel at $T=1.4$~K
($B_{c2}=2.95$~T) and with $I = I_{01}=300,600,800$~nA. The lines
(left-hand scale) correspond to the local response $V_{89}/I_{01}$,
and the symbols (right-hand scale) to $V_{nl}/I = [ V_{45}(I_{01}) -
V_{45}(0)] / I_{01}$. As expected from previous
investigations\cite{weffi} of $V(I)$ of $a$-Nb$_{0.7}$Ge$_{0.3}$, 
for the given current density $J \sim 20 - 50$
MA/m$^2$ the local response at low $T$ depends on $I$. Noteworthy,
in contrast to $d=20$~nm samples with a weaker
pinning\cite{holes,weffi} the onset of $V_l$ at a certain $B=B_{dl}$
is essentially independent of $I$. $B_{dl}$ nearly coincides with
$B_d$ for $L=2$ $\mu$m, while $B_d$ for $L=3,5$ $\mu$m is higher for
$\sim 5$ \% (high $T$) to $\sim 20$ \% (low $T$) and mutually
indistinguishable. Contrary to $V_l(I)$, $V_{nl}(I)$ is for the
given range of $I$ linear for all $T$.\cite{comm1} 

\begin{figure}
\includegraphics[width=70mm]{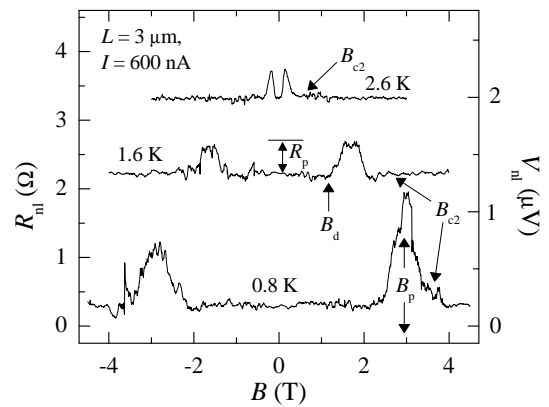}
\caption{$R_{nl}(B)$ (left-hand scale) for an $L=3$ $\mu$m channel
at $T= 0.8, 1.6, 2.6$ K and $I=600$ nA,
with the voltage resolution shown by the right-hand scale.
The curves are offset for clarity.
The meaning of $B_d$, $B_p$ and $R_p$ is indicated by the arrows.}
\end{figure}

In Fig.~2 we show how $R_{nl}=V_{nl}/I$ changes with temperature.
The results refer to an $L=3$ $\mu$m  channel, $I=600$ nA and $T=
0.8, 1.6, 2.6$ K. As in Fig.~1, $R_{nl}=0$ up to $B=B_d$ after which
it displays a relatively broad peak around $B=B_p$, which defines
$R_p = R_{nl}(B_p)$. Close to the $B_{c2}$, $R_{nl}$ drops to zero
again. While the nonlocal resistance in Ref.~\onlinecite{grigorieva}
decreased with decreasing $T$, vanishing at $T/T_c \sim 0.6$, in our
experiment we find a nonmonotonic variation of $R_{nl}(T)$. $R_{nl}$
is finite even at the lowest measurement temperature ($T=0.4$ K,
$T/T_c \approx 0.14$), decreases at low $T$ and increases close to
$T_c$ with increasing $T$ (see Fig.~3(b) later). The $R_{nl}(B)$
traces are nearly symmetric around $B=0$ but, especially at low $T$
and/or high $I$, $R_p$ for $B<0$ and $B>0$ may differ to some
extent, as seen, e.g., in Fig.~1 for $I=800$ nA (1.4 K) and in
Fig.~2 for $T=0.8$ K (600 nA). These differences do not necessarily
appear only because of variations in measurement conditions during
the long $B$ sweeps (e.g., slightly different bath temperature),
they may also originate in the Nernst effect\cite{nernst} due to a
heating in the current-carrying wire.\cite{weffi} However, the
Nernst effect changes sign by reversal of $B$ and can be canceled by
the averaging $R_{nl}(B)=[R_{nl} (+B) + R_{nl}(-B)]/2$. Moreover,
for our samples $R_{nl} (+I) = R_{nl}(-I)$, which rules out a
rectifying effect proposed recently.\cite{vodolazov}

\begin{figure}
\includegraphics[width=65mm]{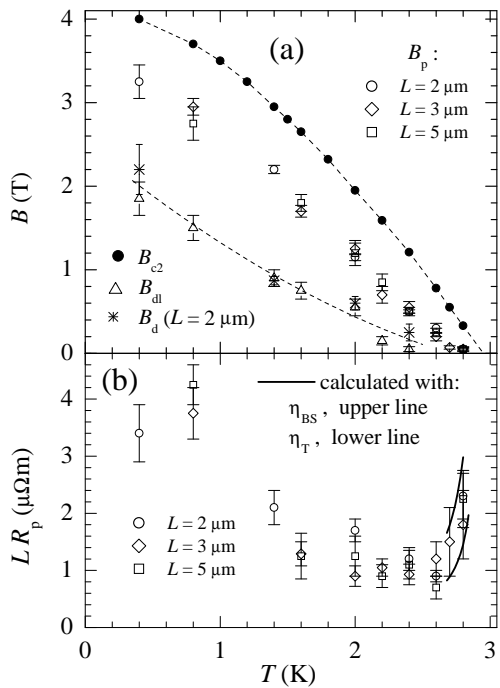}
\caption{Characteristic parameters of $R_{nl}$ for the three channel lengths.
(a) $B_p(T)$ together with $B_{c2}(T)$,
$B_{dl}(T)$ and $B_d(T)$ for the $L=2$ $\mu$m channel. The dashed lines are guides for
the eye. (b) $L R_p$ vs $T$. The solid lines represent
the results of the calculations (using $\eta_{BS}$ and $\eta_T$, as indicated)
explained in the text.}
\end{figure}

Essential information on the nonlocal vortex motion is contained in
the characteristic quantities $B_p$ and $R_p$. In Fig.~3(a) we plot
$B_p(T)$ for all three channel lengths, together with $B_{c2} (T)$,
$B_{dl} (T)$ and $B_d(T)$ for the $L=2$ $\mu$m channel. $B_{dl}$ and
$B_d$ are relatively low and become immeasurably small above 2.6 K
(this applies to $B_d$ for $L=3,5$ $\mu$m as well). It can be seen
that $B_p$ is independent of $L$. On the other hand, $R_p$ does
depend on $L$, which is shown in Fig.~3(b). $R_p L$ vs $T$, plotted
by the symbols, exhibits a reasonably-well-defined scaling behaviour
implying $R_p \propto 1/L$.

We shall now discuss a simple model that leads to a reasonable
explanation of the presented experimental results. This model
assumes equilibrium between the driving force exerted by $I$ on the
vortices in the lower cross and the frictional damping force on the
vortices in the channel, as well as the presence of surface barriers
enforcing a dominating $y$ component of $u$ inside the channel. In a
first step we neglect the pinning and assume that the frictional
force is linear in $u$ with a velocity-independent friction coefficient
$\eta$. The effects of pinning are addressed later.

At a vortex density $n_\phi = B/ \phi_0$, where $\phi_0$ is the
magnetic-flux quantum, in total $n_\phi w^2 $ vortices in the 
lower cross - each experiencing a force $J \phi_0 d$, 
apply a pressure $p= n_\phi \phi_0 I / d$ on the vortices in the channel. The
corresponding pushing force (per unit vortex length) $pw$ is
balanced by the force required to move $ n_\phi Lw$ vortices along
the channel against the frictional damping $\eta u_{nl}$ per vortex.
This gives $u_{nl}= \phi_0 I / \eta L d \propto 1/L$. Using $V_{nl}
= w B u_{nl}$ we find
\begin{equation}
R_{nl} =  \frac{V_{nl}}{I} = \frac{ \phi_0 B w}{L \eta d} \; \;.
\end{equation}
The above result holds if the surface barriers are strong enough to
confine the vortex motion within the channel. As shown
experimentally in Ref.\onlinecite{grigorieva}, this is not satisfied
for large $w$. In any case, surface barriers are essential to
preserve the uniaxial character of the nonlocal vortex motion. An
important source of the surface barriers are the Meissner currents
$J_M$ flowing along the channel edges and providing an
inward-pointing force $F_{in} \propto J_M$. The surface barriers
weaken by approaching $B_{c2}$ irrespectively of their exact origin,
which my explain the fact that $R_{nl}(B)$ does not increase all the
way up to $B=B_{c2}$ before dropping to zero in the normal state.
This restricts the range of the applicability of Eq.~(1) to $B$ not
too close to $B_{c2}$. On the other hand, Eq.~(1) correctly
reproduces\cite{comm2} $R_p \propto 1/L$ and, as we show below,
accounts for $R_p(T)$ quantitatively in conditions of insignificant
pinning.

For $T \geq 2.7$ K the $V_l(I)$ curves are linear beyond any doubt,
implying a negligible pinning and, moreover, $\eta \approx \eta_f$
of the viscous drag in pure flux flow. There are two possible
dissipation mechanisms that determine the flux-flow viscosity
$\eta_f$.\cite{lo} One is related to Joule heating of normal
electrons in vortex cores by the $E$ therein, as described by the
Bardeen-Stephen model giving $\eta_f = \eta_{BS} = \phi_0 B_{c2} /
\rho_n$.\cite{bs} The other approach, proposed by Tinkham,
attributes the dissipation to a loss of the superconducting Gibbs
free-energy density $G_s(B,T)$ as vortices move and cause depairing
and recombination of Cooper pairs.\cite{tinkham} This process is
affected by the time $\tau = (\hbar / \Delta_0 ) [1+ (T/T_c)^2] /
[1-(T/T_c)^2]$ of establishing a superconducting state and results
in $\eta_f = \eta_T = 2 \pi \tau G_s$ ($\Delta_0 \approx 1.76 k_B
T_c$ is the superconducting gap at $T=0$, and $\hbar$ the Planck
constant).\cite{tinkham} For $B$ not too close to $B_{c2}$, i.e. in
the regime of the applicability of Eq.~(1), $G_s$ can be
approximated by the superconducting condensation energy $U_s (T)
\approx B_{c2}^2(T) / 4 \kappa^2 \mu_0$, where $\mu_0 = 4 \pi /
10^7$ H/m.\cite{weffi} Since we know $B_{c2}(T)$ we can determine
both $\eta_{BS}$ and $\eta_T$ for our samples, which permits a
quantitative comparison of the experiment and the model.

At $T \gtrsim 2.7$~K the measured $B_p \approx 60$~mT $\ll B_{c2}$
is fairly constant, so the temperature dependence of the
corresponding $R_p$ is dominated by that of $\eta$. By calculating
$\eta_{BS}$ and $\eta_T$ without any adjustable parameter and using
Eq.~(1) we obtain $L R_p$ shown in Fig.~3(b) by the solid lines, as
indicated. $R_p(T)$ is well reproduced in both cases. 
In particular, the steep increase of $R_p$ near $T_c$ can now be traced back to the
rapid decay of $\eta_f$ close to $T_c$.

So far we have not included any effect of pinning into our analysis,
which is justified only in a narrow $T$ range close to $T_c$. At
lower temperatures the use of $\eta_f$ and the measured values of
$B_p$ in Eq.~(1) results in a rapid increase of $R_{nl}$ far above the
observed values. In addition, nonlinearities in $V_l(I)$ are
observed below 2.7~K, indicating that the pinning is no longer
marginal. However, the fact that the linearity of $V_{nl}(I)$ and
$R_p \propto 1/L$ are preserved in this regime brings up a
possibility of extending our model to lower temperatures. This is
achievable if even in the presence of pinning $p$ remains
proportional to $I$ and $\eta$ independent of $u$. Below we argue
that these properties are consistent with a plausible picture of the
distribution of the pinning force in our samples.

We recall that the onset of $V_l$ at $B_{dl}$ does not depend on
$I$, which suggests a critical vortex density for triggering the
dissipation. This implies that for $B < B_{dl}$ the flux is first
trapped in the most-strong-pinning regions of the
sample.\cite{matsuda} When these are saturated at $B=B_{dl}$
vortices enter the lower-pinning regions in between - where they can
move more easily. Only after this saturation the dissipation starts
by the vortex motion along "vortex rivers" between the
strong-pinning sites/areas. In the current-carrying wire vortices in
the "rivers" shear plastically with the edges, thus enhancing the
driving force above $J \phi_0$ in depinning the vortices. This
process may result in a nonlinear $V_l (I)$ such as that observed in
Fig.~1 but cannot propagate significantly into the channel, where
the external driving current decays exponentially. The pushing force
is thus conveyed along the "rivers" set by the equilibrium pinning
properties, and the fraction of the contributing vortices is $f \sim
(B - B_{dl})/B$. The vortex pressure is therefore reduced by the
same factor $f$.

The frictional damping of the vortex motion in "vortex rivers" has
been investigated in samples with artificial easy-flow channels
embedded in a strong-pinning medium.\cite{droese} Because of the
random pinning landscape in our samples no commensurability effects
between the moving and immobile vortex regions are expected and the
vortex velocity in the "rivers" should respond linearly to the applied
force. This may explain the observed linearity of $V_{nl}(I)$, while
the magnitude of $R_{nl}$ is reduced by the 
ratio $f$ that accounts for the number of vortices in the weak-pinning 
regions. Although we lack a manageable model for
calculating $\eta$ including the effects of pinning, the increase of
$R_p$ as $T$ is lowered can be explained at least qualitatively as a
consequence of $B_p(T)$ growing faster than $\eta(T)$.

To conclude, we have investigated a transversal flux-transformer
effect, which manifests itself in a nonlocal flow of vortices in a
narrow superconducting channel driven by a remote dc current. In our
low-pinning $a$-Nb$_{0.7}$Ge$_{0.3}$ the effect appears in more than half
of the superconducting phase diagram, i.e. everywhere where the
vortices can be moved easily enough to induce dissipation at very
low currents. The effect is two orders of magnitude larger than in
previous studies,\cite{grigorieva} and we observe a nonmonotonic
variation of the maximal nonlocal resistance with temperature, which
can be explained by the interplay of vortex density and vortex
viscosity. Close to $T_c$ the data are in quantitative agreement
with a simple model which is based on the assumption that the
vortices behave like a weakly compressible fluid confined
to the superconducting channel by surface barriers.

We thank E.~H.~Brandt and C.~Morais Smith for helpful discussions.
This work has been financially supported by the DFG within GRK 638
and SFB 631 and by the Croatian MZOS under
project No. 119262.


\begin{thebibliography}{99}
\bibitem{comm2} The 1-$\mu$m-long tails of the crosses, outside the channel area, do
                not add to the damping force. $F_{in}$ pushes vortices inwards
                along the $y$ direction as well, so the contributions of the tails cancel out and $L$ is the proper                                         length for the vortex-block size.
\end{thebibliography}
\end{document}